\documentclass[10pt,twocolumn,a4paper]{article}
\usepackage[left=20mm, right=15mm, top=15mm, bottom=15mm]{geometry}
\usepackage{graphicx,amsmath,amsfonts,amssymb}
\usepackage{flushend}
\usepackage{indentfirst}
\usepackage{epstopdf}
\usepackage{float}

\graphicspath{{./figure/}}
\renewcommand{\Re}{\mbox{Re}}
\begin{document}

\title{\huge \textbf{Application of a Revised
Moli\`{e}re Theory to  the Description of the Landau--Pomeranchuk
Effect}}

\date{31 December 2013}

\twocolumn[
\begin{@twocolumnfalse}
\maketitle

\author\textbf{~~~~~~~~~~~~~~~~~~~~~~~~~~~~~~~~~~~~~Hrach
Torosyan}$^{1}$,
{\textbf{Olga Voskresenskaya}$^{2,*}$\\\\[.5cm]
\footnotesize$ ~~~~~^{1}$Laboratory of Nuclear Problems, Joint Institute
for Nuclear Research, Joliot--Curie 6, 141980, Dubna, Moscow region, Russia\\
\footnotesize $^{2}$ Laboratory of Information Technologies, Joint
Institute for Nuclear Research, Joliot--Curie 6, 141980, Dubna, Moscow region, Russia\\
\footnotesize $~~~~~~~~~~~~~~~~~~~~~~~~~~~~~~~~~~~~~~~~~~~~~~~~~~~~~~~~~~~~~~^{*}$Corresponding Author: voskr@jinr.ru}\\\\\\

\end{@twocolumnfalse}
]

\noindent \textbf{\large{Abstract}} \hspace{2pt} Using the Coulomb
corrections to some important parameters of a revised Moli\`{e}re
multiple scattering  theory, we have obtained analytically and
numerically the Coulomb corrections to the quantities of the Migdal
theory of the Landau--Pomeranchuk (LPM) effect for sufficiently
thick targets. We showed that the Coulomb correction to the spectral
bremsstrahlung rate of this theory allows  completely eliminating
the discrepancy between the theory and experiment at least
for high Z experimental targets.\\

\noindent \textbf{\large{Keywords}} \hspace{2pt}
Landau--Pomeranchuk--Migdal effect, multiple scattering, Coulomb corrections\\

\noindent\hrulefill

\section{\Large{Introduction}}

The theory of the multiple scattering of charged particles has been
treated by several authors [1--7]. However, the most widely used at
present is the multiple scattering theory of Moli\`{e}re \cite{C-7}
whose results are employed nowadays in most of the transport codes.
It is of interest for numerous applications related to particle
transport in matter; and it also presents the most used tool for
taking into account the multiple scattering effects in experimental
data processing. The DIRAC experiment \cite{Dirac05} like many
others \cite{stand} (the MuScat \cite{others1}, MUCOOL
\cite{others2} experiments, etc.) meets the problem of the excluding
of multiple scattering effects in matter from obtained data. The
standard theory of multiple scattering \cite{Dirac05,stand,others1},
proposed by Moli\`ere \cite{C-7} and Fano \cite{Fano}, and some its
modifications \cite{others1,Strig,MIFI} are used for this aim.

As the Moli\`{e}re theory is currently used roughly for $10-300$ GeV
electron beams, the role of the high-energy corrections to the
parameters of this theory becomes significant. Of especial
importance is the Coulomb correction to the screening angular
parameter,  as this parameter also enters into other important
quantities of the Moli\`{e}re theory.

Landau and Pomeranchuk were the first to show \cite{C-2} that
multiplicity of electron scattering processes on atomic nuclei in an
amorphous medium results in the suppression of soft bremsstrahlung.
The quantitative theory of this phenomenon was created by Migdal
\cite{C-4,C-4.5}\footnote{See also \cite{C-5} accounting the edge
effects. Let us notice that Moli\`{e}re's theory was not applied to
the description of the LPM effect in these works. The previous
results of the multiple scattering theory \cite{Rossi} were used
here.}. Therefore, it received the name Landau--Pomeranchuk--Migdal
(LPM) effect.

The analogous effects are possible also at coherent radiation of
relativistic electrons and positrons in a crystalline medium
\cite{Sh1}, in cosmic-ray physics  \cite{Klein1} (e.g. in
applications motivated by extremely high energy IceCubes
neutrino-induced showers with energies above 1 PeV \cite{Klein2}).
Effects of this kind should manifest themselves in scattering of
protons on the nuclei, what has recently been shown in Groning by
the AGOR collaboration \cite{C-1.2}, at penetration of quarks
 and partons through the nuclear matter at the RHIC and LHC energies \cite{KST}.
The QCD analogue of the LPM effect was examined in \cite{Gyu}; a
possibility studying the LPM effect in oriented crystal at GeV
energy was analyzed in \cite{Kryst}. Theoretically, an analogue of
the LPM effect was considered for nucleon-nucleon collisions in the
neutron stars and supernovae \cite{star}, and also in relativistic
plasmas \cite{Plasma}.

The results of a series of experiments at the SLAC
\cite{C-1.1.1,C-1.1.2,Klein3} and CERN-SPS  \cite{C-1.3,C-1.4}
accelerators on detection of the Landau--Pomeranchuk effect
confirmed the basic qualitative conclusion that multiple scattering
of ultrarelativistic charged particles in matter leads to
suppression of their bremsstrah\-lung in the soft part of the
spectrum. However, attempts to quantitatively describe the
experimental data \cite{C-1.1.1} faced an unexpected difficulty. For
achieving satisfactory agreement of data with theory \cite{C-4} the
authors \cite{C-1.1.1} had to multiply the results of their
calculations in the Born approximation by a normalization factor $R$
equal to $0.94\pm 0.01\pm 0.032$, which had no reasonable
explanation.

The alternate calculations \cite{C-3,Cont} gave a similar result
despite different computational basis \cite{C-1.1.1}. The
theoretical predictions are agreement with the spectrum of photon
bremsstrahlung measured for 25 GeV electron beam and
$0.7$--$6.0\%L_{\scriptscriptstyle R}$\footnote{$L_{\scriptscriptstyle \mathrm{R}}$
is the radiation length of the target material.} gold targets over the range
30$<\omega<$500 MeV of the emitted photon frequency $\omega$ only
within a normalization factor $0.94$ \cite{C-1.1.1} -- $0.93$
\cite{C-3}. The origin of the above small but significant
disagreement between data and theory needs to be better understand
\cite{C-1.1.2}.

In  \cite{Cont,Operat1,Z99} the multiphoton effects was taken into
account, and  a comparison with SLAC E-146 data was carried out.
Nevertheless, the problem of normalization remained and is still not
clear. The other authors, except \cite{C-3,Z99}, do not discuss
this normalization problem \cite{Klein3}.

The aim of this work is to show that the  discussed discrepancy can
be explained at least for high Z
targets if the corrections to the results of the Born approximation
are appropriately considered on the basis
of a revised version of the Moli\`{e}re multiple scattering theory
\cite{C-11.1,C-11.2}.

The paper is organized as follows. In Section 2 we consider the
basic formulae of the conventional \cite{C-7} and a revised
\cite{C-11.1,C-11.2} small-angle Moli\`{e}re multiple scattering
theory. We also calculate
the Coulomb corrections to some important parameters of the
Moli\`{e}re theory for the target materials used in \cite{C-1.1.2}.
Then, in Section 3 we present the results of the Migdal LPM effect
theory for sufficiently tick targets and obtain the analytical and
numerical results for the Coulomb corrections to the quantities of
this theory in regimes of the large and the small LPM suppression
based on the Coulomb corrections found in Section 2. Additionally,
we obtain numerical results for Coulomb corrections to the
asymptotes of the spectral radiation rate within the LPM theory
analogue for a thin target \cite{Shul'ga,Fomin}. Finally, in Section 4
we briefly sum up our results.

\section{\Large{Moli\`{e}re's multiple scattering theory}}

\subsection{\normalsize \textbf{Conventional Moli\`{e}re's theory}}

Let $w_{\scriptscriptstyle M}(\vartheta,L)$ be a spatial-angle
particle distribution function in a homogenous medium, and
$\boldsymbol{\vartheta}$ is a two-dimensional particle scattering
angle in the plane orthogonal to the incident particle direction.
For small-angle approximation $\vartheta\ll 1$ ($\sin\vartheta \sim
\vartheta$), the above distribution function is the number of
particles scattered in the angular interval $d\vartheta$ after
traveling through the target of thickness $L$. In the notation of
Moli\`{e}re, it reads
\begin{equation}\label{14}
w_{\scriptscriptstyle M}(\vartheta,L)=
\int\limits_0^\infty J_0(\vartheta
\eta)\exp[-n^{}_{0}L\cdot\nu(\eta)] \eta\, d\eta\ ,
\end{equation}
where
\begin{equation}
\label{15} \nu(\eta)=2\pi \int \limits_0^\infty
\sigma_0(\boldsymbol{\theta})[1-J_0(\theta \eta)]\boldsymbol{\theta}
d\boldsymbol{\theta}\ ,
\end{equation}
$J_0(\vartheta \eta)$ is the Bessel function, $n_0$ denotes the
number density, $\sigma _0(\boldsymbol{\theta})$ presents the Born
differential scattering cross-section, and
$\boldsymbol{\theta}=\boldsymbol{\vartheta}^{\prime}-\boldsymbol{\vartheta}$.

The function (\ref{14}) satisfies the well-known  Boltzmann
transport equation, written here with the small angle approximation
\begin{equation*}\label{Boltzmann}
\frac{\partial w(\vartheta,L)}{\partial
L}=-n_0\,w_{\scriptscriptstyle
M}(\vartheta,L)\int\sigma_0(\boldsymbol{\theta})d^2\boldsymbol{\theta}
\end{equation*}
\begin{equation*}
+ n_0\int w_{\scriptscriptstyle
M}(\boldsymbol{\vartheta}+\boldsymbol{\theta},L)\sigma_0(\boldsymbol{\theta})
d^2\boldsymbol{\theta}\end{equation*}
\begin{equation}
=n_0\int \left[w_{\scriptscriptstyle
M}(\boldsymbol{\vartheta}+\boldsymbol{\theta},L)-w_{\scriptscriptstyle
M}(\vartheta,L)\right]\sigma_0(\boldsymbol{\theta})
d^2\boldsymbol{\theta} \ .
\end{equation}

The Gaussian particle distribution function used in the Migdal LPM
effect theory, which differs from (\ref{14}), can be derived from
the Boltzmann transport equation by the Fokker--Plank method
\cite{C-9}.

One of the most important results of the Moli\`{e}re theory is that
the scattering is described by a single parameter, the so-called
screening angle ($\theta_a$ or $\theta_a^{\,\prime}$)
\begin{equation}\theta_a^{\,\prime}=\sqrt{1.167}\,\theta_a=
\left[\exp\left(C_{\scriptscriptstyle
\mathrm{E}}-0.5\right)\right]\theta_a\approx1.080\,\theta_a\ ,
\end{equation}\\
where $C_{\scriptscriptstyle E}=0.577\ldots$~ is the Euler constant.

More precisely, the angular distribution depends only on the
logarithmic ratio $b$
\begin{equation}\label{b} b=\ln \left(\frac{\theta_c}{\theta_a^{\,\prime}} \right)^2\equiv\ln
\left(\frac{\theta_c}{\theta_a} \right)^2+1-2C_{\scriptscriptstyle
E} \end{equation}\\
of the characteristic angle $\theta_c$ describing the foil thickness
\begin{equation}\label{char} \theta_c^2=4\pi n^{}_0L\left(\frac{Z\alpha}{\beta p}
\right)^2,\quad p=mv\ , \end{equation}\\
to the screening angle $\theta_a^{\,\prime}$, which characterizes
the scattering atom.

In order to obtain a result valid for large angles,  Moli\`{e}re
defined a new parameter $B$ by the transcendental equation
\begin{equation}\label{B} B-\ln B=b\ . \end{equation}\\
The angular distribution function can then be written as
\begin{equation*} w_{\scriptscriptstyle M}(\vartheta,B) =\frac{1}{
\overline{\vartheta^2}}\int\limits_0^{\infty}y dy J_0 (\vartheta
y)e^{-y^2/4}\end{equation*}
\begin{equation}\label{exp2}
\times\exp\left[\frac{y^2}{4B}\ln\left(\frac{y^2}{4}\right)\right],\quad
y=\theta_c\eta\ .
\end{equation}\\

The Moli\`{e}re  expansion method is to consider the term
$y^2\ln(y^2/4)/4B$ as a small parameter.  Then, the angular
distribution function  is expanded in a power series in $1/B$
\begin{equation}\label{power} w_{\scriptscriptstyle
M}(\vartheta,L)=\sum\limits_{n=0}^{\infty}\frac{1}{n!}\frac{1}{B^n}w_n(\vartheta,L)\
,
\end{equation}\\
in which
\begin{equation*} w_n(\vartheta,L) =
\frac{1}{\overline{\vartheta^{\,2}}}\int\limits_0^{\infty}y dy J_0
\left(\frac{\vartheta}{\overline{\vartheta}}\, y\right)
e^{-y^2/4}
\end{equation*}
\begin{equation}
\times\left[\frac{y^2}{4}\ln\left(\frac{y^2}{4}\right)\right]^n\ ,
\end{equation}
\begin{equation}\label{vartheta2}\overline{\vartheta^{\,2}}=\theta_c^2B=4\pi
n^{}_0L\left(\frac{Z\alpha}{\beta p} \right)^2B(L)\ .
\end{equation}\\
This method is valid for $B\geq 4.5$ and
$\overline{\vartheta^{\,2}}<1$.

The first function $w^{}_0(\vartheta,L)$ has a simple analytical
form
\begin{equation}\label{W_0} w^{}_0(\vartheta,L)=\frac{2}{\overline{
\vartheta^{\,2}}}\exp\left(\!\!-\frac{\vartheta^2}{\overline{\vartheta^{\,2}}}\right),
\end{equation}
\begin{equation}\overline{\vartheta^{\,2}} \mathop{\sim}\limits_{\;\;L\,\to \,\infty}\;
\frac{L}{L_{\scriptscriptstyle R}}\ln
\left(\frac{L}{L_{\scriptscriptstyle R}}\right)\ , \end{equation}\\
where For small angles, i.e. $\vartheta/\overline{\vartheta}=
\vartheta/(\theta_c\sqrt{B})$ less than about 2, the Gaussian
\eqref{W_0} is the dominant term.  In this region,
$w_1(\vartheta,L)$ is in general less than $w_0(\vartheta,L)$, so
that the correction to the Gaussian is of order of $1/B$, i.e. about
$10\%$.

A good approximate representation of the distribution at any angle
is
\begin{equation}\label{2order} w_{\scriptscriptstyle M}(\vartheta,L) \approx w_0(\vartheta,L)+
\frac{1}{B}w_1(\vartheta,L)\end{equation}
with
\begin{equation*}w_1(\vartheta,L)=
\frac{1}{\overline{\vartheta^{\,2}}}\int\limits_0^{\infty}y dy J_0
\left(\frac{\vartheta}{\sqrt{\overline{\vartheta^2}}}\, y\right)
e^{-y^2/4}\end{equation*}
\begin{equation}\label{W_1}
\times\left[\frac{y^2}{4}\ln\left(\frac{y^2}{4}\right)\right]\ .
\end{equation}\\
This approximation was applied  by authors of  \cite{Fomin} to the
analysis of data \cite{C-1.1.1,C-1.1.2} over the region $\omega<30$
MeV that will be shown in Section 3.

As show the classical works of Moli\`{e}re \cite{C-7}, the quantity
(\ref{15})  can be represented in the area of the important $\eta$
values  $0\leq \eta\leq 1/\theta_c$ as
\begin{equation}\label{16} \nu(\eta)=-4\pi\Bigg(\frac{Z\alpha}{\beta p}\Bigg)^2
\eta^2\,\left[
\ln\left(\frac{\eta\,\theta_a}{2}\right)+C_{\scriptscriptstyle
E}-\frac{1}{2}\right]\ ,
\end{equation}\\
where the screening angle $\theta_a$ depends both on the screening
properties of the atom and on the $\sigma_0(\boldsymbol{\theta})$
approximation used for its calculation.

Using the Thomas--Fermi model of  the atom and an interpolation
scheme, Moli\`{e}re obtained $\theta_a$ for the cases where
$\sigma_0(\boldsymbol{\theta})$ is calculated within the Born and
quasiclassical approximations:
\begin{equation}\label{17}
\theta_a^{\scriptscriptstyle B}=1.20\cdot\alpha\cdot Z^{1/3}\ ,
\end{equation}\\
\begin{equation}\label{18}
\theta_a^{\scriptscriptstyle M}=\theta_a^{\scriptscriptstyle B}\sqrt{1+3.34\cdot
(Z\alpha/\beta)^2}\ .
\end{equation}\\
Here, $Z$ is the nuclear charge number of the target atom,
$\alpha=1/137$ is the fine structure constant, and $\beta=v/c$ is
the velocity of a projectile in units of the velocity of light.

The latter result (\ref{18}) is only approximate (see critical
remarks on its derivation in \cite{C-9}). Below we will present
exact analytical and numerical results for the screening angle
and some other parameters of the Moli\`{e}re theory.

\subsection{\normalsize\textbf{Revised multiple scattering theory of Moli\`{e}re}}

Very recently, it has been shown \cite{C-11.2} that for any model of
the atom the following rigorous relation determining  the screening
angular parameter $\theta^\prime_a$ is valid:
\begin{equation}\label{Re}
\ln(\theta^\prime_a)=\ln(\theta^\prime_a)^{\scriptscriptstyle B} +
\Re\left[ \psi (1+iZ\alpha/\beta)\right]+C_{\scriptscriptstyle E}
\end{equation}\\
or, equivalently,
\begin{equation}\label{basres}
\Delta_{\scriptscriptstyle
CC}[\ln\big(\theta_a^{\,\prime}\big)]\equiv
\ln(\theta^\prime_a)-\ln(\theta^\prime_a)^{\scriptscriptstyle B}
=f(Z\alpha/\beta)\ ,
\end{equation}\\
where $\Delta_{\scriptscriptstyle CC}$ is the Coulomb correction to
the Born result, $\psi$ is the logarithmic derivative of the gamma
function $\Gamma$, and $f(Z\alpha/\beta)$ is an universal function
of the Born parameter $\xi=Z\alpha/\beta$, which is also known as
the Bethe--Maximon function:
\begin{equation}\label{summa} f(\xi)=\xi^2\sum_{n=1}^\infty\frac{1}{n(n^2+\xi^2)}\ .
\end{equation}\\

To compare the approximate  Moli\`{e}re result (\ref{18}) with the
exact one (\ref{basres}), we first present (\ref{18}) in the form
\begin{equation}
\label{del2}\delta^{}_{\scriptscriptstyle
 M}\left[\theta_a\right]\equiv\frac{\theta^{\scriptscriptstyle M}_a-
 \theta_a^{\scriptscriptstyle B}}
{\theta_a^{\scriptscriptstyle B}}=\sqrt{1+3.34\, \xi^2}-1
\end{equation}\\
and also rewrite (\ref{basres}) as follows:
\begin{equation*}
\delta_{\scriptscriptstyle
CC}\left[\theta_a\right]\equiv\frac{\theta_a-
\theta_a^{\scriptscriptstyle B}}{\theta_a^{\scriptscriptstyle
B}}=\frac{\theta_a^{\,\prime}-
\big(\theta_a^{\,\prime}\big)^{\scriptscriptstyle B}}
{\big(\theta_a^{\,\prime}\big)^{\scriptscriptstyle B}}
\end{equation*}
\begin{equation}\label{del0}
= \exp\left[f\left(\xi\right)\right]-1\ .
\end{equation}\\
Then we get:
\begin{equation}\label{corr3}\delta_{\scriptscriptstyle
CCM}\left[\delta_{\scriptscriptstyle CC}\right]\equiv
\frac{\delta^{}_{\scriptscriptstyle
CC}\left[\theta_a\right]-\delta^{}_{\scriptscriptstyle
M}\left[\theta_a\right]}{\delta^{}_{\scriptscriptstyle
 M}\left[\theta_a\right]}=\frac{\Delta_{\scriptscriptstyle
CCM}\left[\delta_{\scriptscriptstyle
CC}\right]}{\delta^{}_{\scriptscriptstyle M}\left[\theta_a\right]}
\,.
\end{equation}\\

In order to obtain the relative difference between the  approximate
$\theta_a^{\scriptscriptstyle M}$ and exact $\theta_a$ results for
the screening angle
\begin{eqnarray}\label{angle}
\delta_{\scriptscriptstyle
CCM}[\theta_a]&\equiv&\frac{\theta_a-
\theta_a^{\scriptscriptstyle M}}
{\theta_a^{\scriptscriptstyle M}}=\frac{\theta_a}
{\theta_a^{\scriptscriptstyle M}}-1\\
\label{angle-1} &=& R_{\scriptscriptstyle CCM}[\theta_a]-1\
,
\end{eqnarray}\\
we rewrite (\ref{del2}) and (\ref{del0}) in the following
form
\begin{equation}
\label{folform}
\delta_{\scriptscriptstyle CC}[\theta_a]+1=\frac{\theta_a}
{\theta_a^{\scriptscriptstyle B}}\ ,\qquad
\delta^{}_{\scriptscriptstyle
M}[\theta_a]+1=\frac{\theta^{\scriptscriptstyle M}_a}
{\theta_a^{\scriptscriptstyle B}}\end{equation}\\
and obtain for the ratio $R_{\scriptscriptstyle
CCM}[\theta_a]$ the expression
\begin{eqnarray}
\label{rat} R_{\scriptscriptstyle
CCM}[\theta_a]&\equiv&\frac{\theta_a}
{\theta_a^{\scriptscriptstyle
M}}=\frac{\delta_{\scriptscriptstyle CC}
[\theta_a]+1}{\delta^{}_{\scriptscriptstyle
M}[\theta_a]+1}
\end{eqnarray}
\begin{equation}
=\delta_{\scriptscriptstyle CCM}[\theta_a]+1\ .
\end{equation}\\
We can also represent the relative difference (\ref{angle}) by the
equation
\begin{equation}\label{reldif}
\delta_{\scriptscriptstyle
CCM}[\theta_a]=\frac{\Delta_{\scriptscriptstyle
CCM}[\delta_{\scriptscriptstyle CC}]}{\delta^{}_{\scriptscriptstyle
M}[\theta_a]+1}\ .
\end{equation}

\begin{center} {\bf Table 1.}
Numerical results for the relative corrections (\ref{del2}),
(\ref{del0}), relative differences (\ref{corr3}), (\ref{angle-1}),
and the ratio (\ref{rat}) in the range of nuclear charge $73\leq
 Z \leq 92$.

\bigskip

\begin{tabular}{cccccc}
\hline \\[-2mm]
Z&$\delta_{\scriptscriptstyle
M}[\theta_a]$&$\!\!\delta_{\scriptscriptstyle CC}[\theta_a]\!\!$&
$\!\!\delta_{\scriptscriptstyle CCM}[\delta_{\scriptscriptstyle
CC}]$\!\!&$\delta_{\scriptscriptstyle
CCM}[\theta_a]$&$R_{\scriptscriptstyle
CCM}[\theta_a]$ \\[.2cm]
\hline\\[-3mm]
73&0.396&0.318&$-0.198$&$-0.056$&0.944\\
74&0.404&0.325&$-0.196$&$-0.056$&0.943\\
78&0.443&0.359&$-0.189$&$-0.058$&0.942\\
79&0.452&0.367&$-0.188$&$-0.059$&0.941\\
82&0.482&0.393&$-0.185$&$-0.060$&0.940\\
92&0.583&0.485&$-0.169$&$-0.062$&0.938\\[.2cm]
\hline
\end{tabular}
\end{center}

\bigskip

For some high Z targets used in \cite{C-1.1.2} and $\beta=1$, we
obtain the following values of the relative Moli\`{e}re
$\delta_{\scriptscriptstyle M}[\theta_a]$ (\ref{del2}) and Coulomb
$\delta_{\scriptscriptstyle CC}[\theta_a]$ (\ref{del0}) corrections
and also the sizes of the difference $\Delta_{\scriptscriptstyle
CCM}[\delta_{\scriptscriptstyle CC}]$ and relative differences
$\delta_{\scriptscriptstyle CCM}[\delta_{\scriptscriptstyle CC}]$
(\ref{corr3}), $\delta_{\scriptscriptstyle CCM}[\theta_a]$
(\ref{angle-1}) as well as the ratio $R_{\scriptscriptstyle
CCM}[\theta_a]$ (\ref{rat}) (Table 1, Figure 1).

From the Table 1 it is evident that the Coulomb correction
$\delta_{\scriptscriptstyle CC}[\theta_a]$ has a large
value, which ranges from around $30\%$ for $Z\sim 70$ up to $50\%$
for $Z\sim 90$. The relative difference between the approximate and
exact results for this Coulomb correction varies from 17 up to
$20\%$ over the range $73\leq Z\leq 92$.

The relative difference $\delta_{\scriptscriptstyle CCM}[\theta_a]$
between the approximate $\theta^{\scriptscriptstyle M}_a$ and exact
$\theta_a$ results for the screening angle as well as
$R_{\scriptscriptstyle
CCM}[\theta_a]=\theta_a/\theta^{\scriptscriptstyle M}_a$ value does
not vary significantly from one target material to another. Their
sizes are $5.86\pm 0.22\%$ for $-\delta_{\scriptscriptstyle
CCM}[\theta_a]$ and $0.941\pm 0.002$ for $R_{\scriptscriptstyle
CCM}[\theta_a]$ over the Z range studied. It is interesting that the
latter value coincides with the normalization constant $R=0.94\pm
0.01$ found in \cite{C-1.1.1}.

We show further that the above discrepancy between theory of the LPM
effect and experiment \cite{C-1.1.1,C-1.1.2,C-3} can be completely
eliminated for heavy target elements on the basis of the Coulomb
corrections to the screening angular parameter. For this purpose, we
calculate also some additional Coulomb corrections to other
important parameters of the Moli\`{e}re theory. Inserting (\ref{b})
into (\ref{B}) and differentiating the latter, we arrive at
\begin{equation}\label{limcorrec1}
\Delta_{\scriptscriptstyle CC}[b]=-f(\xi)
=\left(1-\frac{1}{B^{\scriptscriptstyle
B}}\right)\cdot\Delta_{\scriptscriptstyle CC}[B]\ .
\end{equation}
So $\Delta_{\scriptscriptstyle CC}[B]$ becomes
\begin{equation}\label{limco1}
\Delta_{\scriptscriptstyle
CC}[B]=\frac{f(\xi)}{1/B^{\scriptscriptstyle B}-1}\ .
\end{equation}\\
Accounting $\overline{\vartheta^2}=\theta_c^2B$ (\ref{vartheta2}),
we get
\begin{equation}\label{CCvarthet2}
\Delta_{\scriptscriptstyle
CC}\left[\overline{\vartheta^2}\right]\equiv
\overline{\vartheta^2}-\left(\overline{\vartheta^2}\right)^{\scriptscriptstyle
B}=\theta_c^2 \cdot\Delta_{\scriptscriptstyle CC}\left[B\right]\ .
\end{equation}\\
Finally, the relative Coulomb corrections can be represented as
\begin{equation}\label{rel}
\delta_{\scriptscriptstyle
CC}\left[\overline{\vartheta^2}\right]=\delta_{\scriptscriptstyle
CC}\left[B\right] =\frac{f(\xi)}{1-B^{\scriptscriptstyle B}}\ .
\end{equation}
The Z dependence of the corrections (\ref{limcorrec1}),
(\ref{limco1}), and (\ref{rel}) is presented in Table 2 (see also
Figure 1).

\begin{center} {\bf Table 2.}
The Coulomb correction (\ref{limcorrec1}), (\ref{limco1}), and
(\ref{rel}) to the parameters of the Moli\`{e}re theory for
$B^{\scriptscriptstyle B}=8.46$ and $\beta=1$.
\end{center}

\vspace{.25mm}

\begin{center}
\begin{tabular}{cccccc}
\hline \\[-2mm]
M&Z&$\Delta_{\scriptscriptstyle CC}[b]$&$\Delta_{\scriptscriptstyle
CC}[B]$&$\delta_{\scriptscriptstyle
CC}\left[B\right]$&$\delta_{\scriptscriptstyle
CC}\left[\overline{\vartheta^2}\right]$\\[.2cm]
\hline\\[-4mm]
Al&13&$-0.0107$&$-0.0121$  &$-0.0014$&$-0.0014$\\
Fe&26&$-0.0420$&$-0.0476$  &$-0.0056$&$-0.0056$ \\
W &74&$-0.2813$&$-0.3190$  &$-0.0377$&$-0.0377$ \\
Au&79&$-0.3125$&$-0.3545$  &$-0.0419$&$-0.0419$\\
Pb&82&$-0.3316$&$-0.3760$  &$-0.0445$&$-0.0445$\\
 U&92&$-0.3951$&$-0.4481$  &$-0.0530$& $-0.0530$ \\[.2cm]
\hline
\end{tabular}
\end{center}

\begin{figure}[h!]

\begin{center}
\includegraphics[width=1\linewidth]{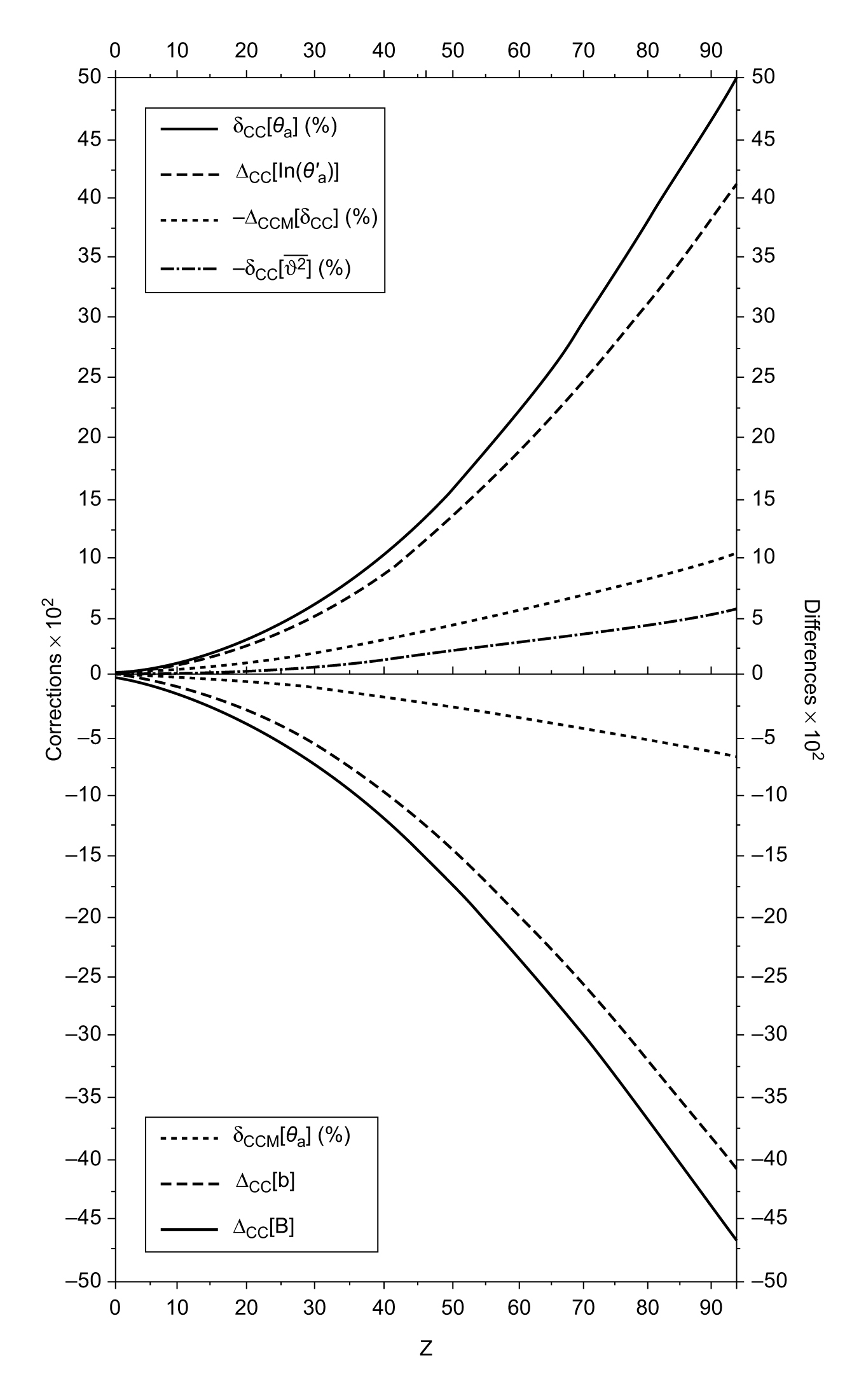}
\caption{The $Z$ dependence of the Coulomb corrections to some
parameters of the Moliere theory and the differences between exact
and approximate results \cite{C-11.2}.\label{}}

\end{center}

\end{figure}
\section{\Large\textbf{Applications of the
Moli\`{e}re theory to  the description of the LPM effect and its
analogue}}

\subsection{\normalsize\textbf{Basic formulae of the Migdal LPM effect
theory for sufficiently thick targets}}

There exist two methods that allow to develop a rigo\-rous
quantitative theory of the Landau--Pomeranchuk effect. It is
Migdal's method of kinetic equation \cite{C-4,C-4.5}  and the method
of functional integration \cite{C-3,Cont,Z99,path,C-6}.
Neglecting numerically small quantum-mechanical corrections, we will adhere to
version of the Landau--Pomeranchuk effect theory developed in
\cite{C-4}.

Simple though quite cumbersome calculations yield the following
formula for the electron spectral bremsstrahlung intensity averaged
over various trajectories of electron motion in an amorphous medium
(hereafter the units  $\hbar = c =1$, $e^2=1/137$ are used):

\begin{equation}\label{migdal}
\left\langle\frac{dI}{d\omega}\right\rangle
=\Phi(s)\left(\frac{dI}{d\omega}\right)_0\ ,
\end{equation}\\
where $(dI/d\omega)_0$ is the spectral bremsstrahlung rate without
accounting for the multiple scattering effects in the radiation,
\begin{equation}\label{w0}
\left(\frac{dI}{d\omega}\right)_0=\frac{2e^2}{3\pi} \gamma^2q\,L\ ,
\end{equation}
\begin{equation}\label{q}
q=\overline{\vartheta^2}/L\ ,
\end{equation}\\
and $\gamma$ is the Lorentz factor of the scattered particle.

The function $\Phi(s)$  accounts for the multiple scattering
influence on the bremsstrahlung rate and reads
\begin{equation}\label{Phi}
\Phi(s)=24s^2\left[\int\limits_{0}^{\infty}
dx\,e^{-2sx}\mbox{cth}(x)\sin(2sx)-\frac{\pi}{4} \right],
\end{equation}
\begin{equation}\label{s2}
s^2=\lambda^2/\overline{\vartheta^2},\quad \lambda^2=\gamma^{-2}\ .
\end{equation}\\
It has simple asymptotes at the small and large values of the
argument:
\begin{equation}\label{as}
\Phi(s) \rightarrow\left\{\begin{array}{cl}
6s,&s\;\rightarrow \;0\;,\\
1,&s\;\rightarrow \infty,\end{array}\right.
\end{equation}
\begin{equation}\label{s}
s=\frac{1}{4\gamma^2}\sqrt{\frac{\omega}{q}}\ .
\end{equation}\\
For $s\ll 1$, the suppression is large, and $\Phi(s)\approx 6s$. The
intensity of radiation in this case is much less, than the
corresponding result of Bethe and Heitler. If $s\geq 1$, the function
$\Phi(s)$ is close to a unit,  and the following approximation is valid
\cite{Sh1}:
\begin{equation}\label{small}
\Phi(s)\approx 1-0.012/s^4\ .
\end{equation}

The formula (\ref{migdal}) is obtained with the logarithmic
accuracy. At $s\gg 1$, (\ref{migdal}) coincides within this accuracy
with the Bethe--Heitler result
\begin{equation}\label{BH}
\left\langle\frac{dI}{d\omega}\right\rangle_{\scriptscriptstyle
 BH} =\frac{L}{L_{\scriptscriptstyle
R}}\left[1+\frac{1}{12\ln\left(183Z^{-1/3}\right)}\right]\ .
\end{equation}\\
If $s\ll 1$, we have the large LPM suppression in comparison with
(\ref{BH}). Let us notice that effect of a medium polarization is
not considered here, i.e. it is assumed that the absolute
permittivity of the medium $\varepsilon(\omega)=1$.

\subsection{\normalsize\textbf{Applying the revised
theory of Moli\`{e}re to the Migdal LPM effect theory}}

Now we obtain analytical and numerical results for the Coulomb
corrections to the quantities of the Migdal LPM effect theory. In
order to derive an analytical expression for the Coulomb correction
to the Born spectral bremsstrahlung rate $(dI/d\omega)_0$, we first
write
\begin{equation*}
\Delta_{\scriptscriptstyle
CC}\left[\left(\frac{dI}{d\omega}\right)_0\right]\equiv
\left(\frac{dI}{d\omega}\right)_0-
\left(\frac{dI}{d\omega}\right)^{\scriptscriptstyle B}_0
\end{equation*}
\begin{equation}\label{CCw0}
=\frac{2e^2}{3\pi} \gamma^2L\cdot\Delta_{\scriptscriptstyle CC}[q]\
 ,
\end{equation}
where
\begin{equation*}\label{CCq}
\Delta_{\scriptscriptstyle CC}[q]\equiv q-q^{\scriptscriptstyle
B}=\frac{1}{L}\cdot\Delta_{\scriptscriptstyle
CC}\left[\overline{\vartheta^2}\right]\ ,
\end{equation*}
\begin{equation*}\label{CCvartheta2}
\Delta_{\scriptscriptstyle
CC}\left[\overline{\vartheta^2}\right]\equiv
\overline{\vartheta^2}-\left(\overline{\vartheta^2}\right)^{\scriptscriptstyle
B}=\theta_c^2 \cdot\Delta_{\scriptscriptstyle CC}\left[B\right]\ ,
\end{equation*}
\begin{equation}\label{limcor1}
\Delta_{\scriptscriptstyle
CC}[B]=\frac{f(\xi)}{1/B^{\scriptscriptstyle B}-1}\ .
\end{equation}\\
In doing so, (\ref{CCw0}) becomes
\begin{equation}
\Delta_{\scriptscriptstyle
CC}\left[\left(\frac{dI}{d\omega}\right)_0\right]
=\frac{2(e\gamma\theta_c)^2}{3\pi\,(1/B^{\scriptscriptstyle
B}-1)}\cdot f(\xi)\ ,
\end{equation}\\
and the relative Coulomb correction reads
\begin{equation*}
\delta_{\scriptscriptstyle
CC}\left[(dI/d\omega)_0\right]=\delta_{\scriptscriptstyle
CC}\left[q\right]=\delta_{\scriptscriptstyle
CC}\left[\overline{\vartheta^2}\right]
\end{equation*}
\begin{equation}\label{migdalparam}
=R_{\scriptscriptstyle
CC}\left[(dI/d\omega)_0\right]-1=\frac{f(\xi)}{1-B^{\scriptscriptstyle
B}}\ .
\end{equation}

Next, in order to obtain the relative Coulomb correction to the
Migdal function $\Phi(s)$, we first derive corresponding correction
to the parameter $s^2$ (\ref{s2}):
\begin{equation}
\Delta_{\scriptscriptstyle CC}\left[s^2
\right]=\frac{\omega}{16\gamma^4}
\left(\frac{1}{q}-\frac{1}{q^{\scriptscriptstyle B}}\right)\ ,
\end{equation}
\begin{equation}\label{CCs2}
\delta_{\scriptscriptstyle CC}\left[s^2
\right]=\frac{q^{\scriptscriptstyle
B}}{q}-1=\frac{\big(\overline{\vartheta^2}\big)^{\scriptscriptstyle
B}}{\overline{\vartheta^2}}-1\ ,
\end{equation}
\begin{equation}
\frac{\big(\overline{\vartheta^2}\big)^{\scriptscriptstyle
B}}{\overline{\vartheta^2}}=\frac{1}{ \delta_{\scriptscriptstyle
CC}\big[\overline{\vartheta^2}\big]+1}\ .
\end{equation}
This leads to the following relative Coulomb correction for $s$
(\ref{s}):
\begin{equation*}
\delta_{\scriptscriptstyle CC}\left[s \right]
=\sqrt{\frac{\big(\overline{\vartheta^2}\big)^{\scriptscriptstyle
B}}{\overline{\vartheta^2}}}-1
\end{equation*}
\begin{equation}
= \frac{1}{\sqrt{\delta_{\scriptscriptstyle
CC}\big[\overline{\vartheta^2}\big]+1}}-1\ .
\end{equation}

For the asymptote $\Phi(s)=6s$ (\ref{as}), we get
\begin{equation}\label{CCPhi}
\delta_{\scriptscriptstyle CC}\left[\Phi(s) \right]
=\delta_{\scriptscriptstyle CC}\left[s \right]
=\frac{1}{\sqrt{R_{\scriptscriptstyle
CC}\left[(dI/d\omega)_0\right]}} -1\ .
\end{equation}

Then the total relative Coulomb correction to  the spectral density
of radiation in this asymptotic case becomes
\begin{equation}\label{sum}
\delta_{\scriptscriptstyle CC}\left[\langle dI/d\omega\rangle
\right]=\delta_{\scriptscriptstyle
CC}\left[(dI/d\omega)_0\right]+\delta_{\scriptscriptstyle
CC}\left[\Phi(s) \right]\ .
\end{equation}

The regime of strong LPM suppression is not reached in the conditions of the
experiment \cite{C-1.1.1,C-1.1.2,Klein3}. Therefore, we will carry
out now calculation for the regime of small LPM suppression
\eqref{small}.

In order to obtain the relative correction
$\delta_{\scriptscriptstyle \mathrm{CC}}\left[\Phi(s) \right] $ in
this regime, we first derive an expression for the Coulomb
correction $\Delta_{\scriptscriptstyle \mathrm{CC}}\left[\Phi(s)
\right]$ to the Migdal function $\Phi(s)$:
\begin{equation*}
\Delta_{\scriptscriptstyle CC}\left[\Phi(s) \right]=0.012
\left(\frac{1}{\left(s^4\right)^{\scriptscriptstyle B}}-
\frac{1}{s^4}\right)=\frac{0.012}{s^4}\,\delta_{\scriptscriptstyle
CC}\left[s^4 \right]\ ,
\end{equation*}
\begin{equation*}
\delta_{\scriptscriptstyle CC}\left[s^4
\right]=\left(\frac{q^{\scriptscriptstyle
B}}{q}\right)^2-1=\left(\frac{\big(\overline{\vartheta^2}\big)^{\scriptscriptstyle
B}}{\overline{\vartheta^2}}\right)^2-1
\end{equation*}
\begin{equation}\label{CCs4}
=\frac{1}{\left(\delta_{\scriptscriptstyle
CC}\big[\overline{\vartheta^2}\big]+1\right)^2}-1
=\frac{1}{\Big(R_{\scriptscriptstyle
CC}\left[(dI/d\omega)_0\right]\Big)^2}-1\ .
\end{equation}

This leads to the following relative Coulomb correction for
$\Phi(s)$ (\ref{small}):
\begin{equation*}
\delta_{\scriptscriptstyle CC}\left[\Phi(s) \right]
=\frac{0.012}{s^4}\,\delta_{\scriptscriptstyle CC}\left[s^4
\right] \cdot\frac{\left(s^4\right)^{\scriptscriptstyle B}
}{\left(s^4\right)^{\scriptscriptstyle B}-0.012}\ .
\end{equation*}
\begin{equation}\label{CCPhismall}
=0.012\,\frac{\delta_{\scriptscriptstyle CC}\left[s^4
\right]}{\delta_{\scriptscriptstyle CC}\left[s^4 \right]+1}
\cdot\frac{1}{\left(s^4\right)^{\scriptscriptstyle
\mathrm{B}}-0.012}\ .
\end{equation}

In Table 3 are listed the values of the relative Coulomb corrections
to the quantities of \eqref{migdal} in the regime of small LPM
suppression (\ref{small}) for some separate $s$ values
from the range $1.0 \leq s\leq \infty$ (e.g., for $s=1.1$ and $s=1.5$).

\begin{center}
{\bf Table 3.} Coulomb corrections  to the quantities of the Migdal
LPM theory, $\delta_{\scriptscriptstyle
CC}\left[(dI/d\omega)_0\right]$ (\ref{migdalparam}),
$\delta_{\scriptscriptstyle CC}\left[\Phi(s) \right]$
(\ref{CCPhismall}), and $\delta_{\scriptscriptstyle CC}\left[\langle
dI/d\omega\rangle\right]$ (\ref{sum}), in the regime of small LPM
suppression for high Z targets of experiment \cite{C-1.1.2}\footnote{For low Z
targets, the E-146 data showed a disagreement with the Migdal LPM
theory predictions. There is a problem of an adequate describe the
photon spectra shape for the low Z targets \cite{C-1.1.2,Klein3}}.
\smallskip

{\bf 1.} at $\beta=1$, $B^{\scriptscriptstyle B}=8.46$, and $s=1.1$

\medskip

\begin{tabular}{ccccc}
\hline \\[-3mm]
Z&$\delta_{\scriptscriptstyle
CC}\left[\left(\frac{dI}{d\omega}\right)_{\scriptscriptstyle
0}\right]$ &$\delta_{\scriptscriptstyle CC}\left[s^4
\right]$&$\delta_{\scriptscriptstyle CC}\left[\Phi(s) \right]
$&$\delta_{\scriptscriptstyle
CC}\left[\left\langle\frac{dI}{d\omega}\right\rangle\right]$\\[.2cm]
\hline\\[-3mm]
79&$-0.0419$&$-0.0896$&$-0.0008$&$-0.0427$\\
82&$-0.0445$&$-0.0953$&$-0.0009$&$-0.0454$\\
92&$-0.0530$&$-0.1149$&$-0.0011$&$-0.0541$\\[2mm]
\hline
\end{tabular}

\medskip

$\bar{\delta}_{\scriptscriptstyle CC}\left[\left\langle
dI/d\omega\right\rangle\right]= -4.74\pm 0.59\%$;\\

\bigskip

{\bf 2.} at $\beta=1$, $B^{\scriptscriptstyle B}=8.46$, and $s=1.5$\\

\medskip

\begin{tabular}{ccccc}
\hline \\[-3mm]
Z&$\delta_{\scriptscriptstyle
CC}\left[\left(\frac{dI}{d\omega}\right)_{\scriptscriptstyle
0}\right]$ &$\delta_{\scriptscriptstyle CC}\left[s^4
\right]$&$\delta_{\scriptscriptstyle CC}\left[\Phi(s) \right]
$&$\delta_{\scriptscriptstyle
CC}\left[\left\langle\frac{dI}{d\omega}\right\rangle\right]$\\[.2cm]
\hline\\[-3mm]
79&$-0.0419$&$-0.0896$&$-0.0002$&$-0.0421$\\
82&$-0.0445$&$-0.0953$&$-0.0002$&$-0.0447$\\
92&$-0.0530$&$-0.1149$&$-0.0003$&$-0.0533$\\[.2cm]
\hline
\end{tabular}

\medskip

$\bar{\delta}_{\scriptscriptstyle CC}\left[\left\langle
dI/d\omega\right\rangle\right]= -4.67\pm 0.57\%$.\\

\end{center}

Table 4 presents the average values of the corrections
$-{\delta}_{\scriptscriptstyle CC}\left[\left\langle
dI/d\omega\right\rangle\right]\,(\%)$ for  separate high Z target
elements and the common average $-\bar{\delta}_{\scriptscriptstyle
CC}\left[\left\langle dI/d\omega\right\rangle\right]\,(\%)$ over
the entire range  $1.0 \leq s\leq \infty$ of the parameter $s$, for
which the regime of small LPM suppression is valid.

\begin{center}
{\bf Table 4.} The dependence of the relative Coulomb correction
$-\delta_{\scriptscriptstyle CC}\left[\langle
dI/d\omega\rangle\right]$ value ($\%$) on the parameter $s$ in the
regime of small LPM suppression for high Z targets, $\beta=1$, and
$B^{\scriptscriptstyle B}=8.46$.

\medskip

\begin{tabular}{cccccccc}
\hline \\[-2mm]
$Z\diagdown s$&1.0 &1.1 &1.2 &1.3&1.5&2.0&$\infty$\\[.2cm]
\hline\\[-3mm]
79&4.32&4.28&4.26&4.24&4.22&4.21&4.19\\
82&4.58&4.54&4.51&4.49&4.47&4.46&4.45\\
92&5.45&5.41&5.36&5.34&5.33&5.31&5.30\\[.2cm]
\hline
\end{tabular}

\medskip
$\delta_{\scriptscriptstyle CC}\left[\left\langle
dI/d\omega\right\rangle\right]=-4.50\pm 0.05\%$ ($Z=82$)\ ,
\smallskip
$\delta_{\scriptscriptstyle CC}\left[\left\langle
dI/d\omega\right\rangle\right] =-5.35\pm 0.06\%$ ($Z=92$)\ ,
\smallskip
 $\bar{\delta}_{\scriptscriptstyle
CC}\left[\left\langle dI/d\omega\right\rangle\right] = -4.70\pm
0.49\%$\ .

\end{center}

It will be seen from Table 4  that the Coulomb corrections
$\delta_{\scriptscriptstyle CC}\left[\langle
dI/d\omega\rangle\right]=-4.50\pm 0.05\%$ ($Z=82$) and
$\delta_{\scriptscriptstyle CC}\left[\langle
dI/d\omega\rangle\right]=-5.35\pm 0.06\%$ ($Z=92$) coincide within
the experimental error with the sizes of the normalization
correction $-4.5\pm 0.2\%$ for $2\%L_{\scriptscriptstyle R}$ lead
target and $-5.6\pm 0.3\%$ for $3\%L_{\scriptscriptstyle R}$ uranium
target, respectively (Table II in \cite{C-1.1.2}).

It is also obvious that the average
$\delta_{\scriptscriptstyle CC}\left[\langle
dI/d\omega\rangle\right]$ value $\bar{\delta}_{\scriptscriptstyle
CC}\left[\left\langle dI/d\omega\right\rangle\right]= -4.70\pm
0.49\%$ excellent agrees with the  weighted average $-4.7\pm 2\%$ of
the normalization correction obtained in \cite{C-1.1.2} for 25 GeV
data\footnote{It becomes $-4.8\pm 3.5\%$ for the 8 GeV data if the
outlying $6\%L_{\scriptscriptstyle R}$ gold target is excluded from
them \cite{C-1.1.2}.}.

We believe that this allows to understand an
origin of the discussed in \cite{C-1.1.1,C-1.1.2} normalization
problem for high Z targets.

\subsection{\normalsize\textbf{Application of  Moli\`{e}re's theory to the
description of the LPM effect analogue for a thin target}}

Experiment \cite{C-1.1.1,C-1.1.2} caused considerable interest and
stimulated development of various approaches to the study of the LPM
effect, including an application of Moli\`{e}re's  results to the
description of an analogue of the LPM effect for a thin layer of
matter \cite{Fomin}\footnote{The authors of \cite{Fomin} neglect the
influence of the medium polarization \cite{TM} on the radiation in
this theory.}.

In \cite{Fomin} it is shown that the region of the emitted photon
frequencies naturally splits into two intervals, $\omega
> \omega_{c}$ and $\omega<\omega_{c}$, in first of which the
LPM effect for sufficiently tick targets takes place, and in the
second, there is its analogue for thin targets. The quantity
$\omega_{c}$ is defined here as $\omega_{c}=2\gamma^2/L$.

Application of the Moli\`{e}re multiple scattering theory to the
analysis of experimental data \cite{C-1.1.1,C-1.1.2} for a thin
target in the second $\omega$ range is based on the use of the
expression for the spatial-angle particle distribution function
(\ref{14}), which satisfies the standard Boltzmann transport
equation for a thin homogenous foil and differs significantly from
the Gaussian particle distri\-bu\-tion of the Migdal LPM effect
theory.

Besides, it determines an another expression for the spectral
radiation rate in the context of the coherent radiation theory
\cite{Fomin}, which reads
\begin{equation}\label{averaged}
\,\left\langle\frac{dI}{d\omega}\right\rangle = \int
w_{\scriptscriptstyle
M}(\vartheta)\frac{dI(\vartheta)}{d\omega}d^2\vartheta\ .
\end{equation}
Here
\begin{equation}\label{nonaveraged}
\frac{dI(\vartheta)}{d\omega}=\frac{2e^2}{\pi}
\left[\frac{2\chi^2+1}{\chi\sqrt{\chi^2+1}}
\ln\left(\chi+\sqrt{\chi^2+1}\right)-1\right]
\end{equation}\\
with $\chi=\gamma\vartheta/2$. The latter expression is valid for
consideration of the particle scattering in both amorphous and
crystalline medium.

The formula (\ref{nonaveraged}) has simple asymptotes at the small
and large values of  parameter $\chi$:
\begin{equation}\label{asymptnon}
\frac{dI(\vartheta)}{d\omega}
=\frac{2e^2}{3\pi}\left\{\begin{array}{cl}
\gamma^2\vartheta^2,&\gamma\vartheta\ll 1\;,\\
3\left[\ln (\gamma^2\vartheta^2)-1\right],&\gamma\vartheta\gg
1\;,\end{array}\right.
\end{equation}\\

Replacing in this formula  $\vartheta^2$  by the average square
value of the scattering angle $\overline{\vartheta^2}$, we arrive at
the following estimates for the average radiation spectral density
value:

\begin{equation}\label{asymptaver}
\left\langle\frac{dI}{d\omega}\right\rangle
=\frac{2e^2}{3\pi}\left\{\begin{array}{cl}
\gamma^2\overline{\vartheta^2},&\gamma^2\overline{\vartheta^2}\ll 1\;,\\
3\left[\ln
(\gamma^2\overline{\vartheta^2})-1\right],&\gamma^2\overline{\vartheta^2}\gg
1\;.\end{array}\right.
\end{equation}\\

In the experiment  \cite{C-1.1.1,C-1.1.2},  the above  frequency
intervals correspond roughly to the following $\omega$ ranges:
$(\omega > \omega_{c})\sim (\omega > 30\,\mbox{MeV})$ and $(\omega <
\omega_{c})\sim (\omega < 30\,\mbox{MeV} )$ for 25 GeV electron beam
and $0.7-6.0\%L_{\scriptscriptstyle R}$ gold targets. Whereas in the
first area the discrepancy between the LPM theory predictions and
data is about 3.2 to 5\% that requires the use of normalization
factor $0.94\pm 0.01\pm 0.032$, in the second area this discrepancy
reaches $\sim 15\%$.

Using the second-order representation of the Moli\`{e}re
distri\-bu\-tion function (\ref{2order}), (\ref{W_1}) for computing
the spectral radiation rate (\ref{averaged}) the authors of
\cite{Fomin} were able to agree satisfactorily theory and 25 GeV and
$0.7\%L_{\scriptscriptstyle R}$ data over the range $\omega<30$ MeV.

This result can be understood by considering the fact that the
correction to the Gaussian first-order representation of the
distribution function $w_{\scriptscriptstyle M}(\vartheta)$ of order
of $1/B^{\scriptscriptstyle B}$ is about 12\% for the used in
calculations value $B^{\scriptscriptstyle B}=8.46$ \cite{Fomin}.

\subsection{\normalsize\textbf{Coulomb corrections in the coherent
radiation theory for a thin target}}

Let us obtain the relative Coulomb correction to the average value
of the spectral density of radiation for two limiting cases
(\ref{asymptaver}).

In the first case $\gamma^2\overline{\vartheta^2}\ll 1$, takin into
account the equality
\begin{equation}
\delta_{\scriptscriptstyle
CC}[\gamma^2\overline{\vartheta^2}]=\delta_{\scriptscriptstyle
CC}[\overline{\vartheta^2}]\ ,
\end{equation}\\
(\ref{migdalparam}),  and (\ref{asymptaver}), we get
\begin{equation}\label{limcorrect1}
\delta_{\scriptscriptstyle CC}\left[\left\langle
\frac{dI}{d\omega}\right\rangle\right]=\delta_{\scriptscriptstyle
CC}\left[\left( \frac{dI}{d\omega}\right)_0\right]
=\frac{f(\xi)}{1-B^{\scriptscriptstyle B}}\ ,
\end{equation}\\
where $B^{\scriptscriptstyle B} \approx 8.46$ in the conditions of
the discussed experiment \cite{Fomin}.

In the second case $\gamma^2\overline{\vartheta^2}\gg 1$, we have
\begin{equation*}
\Delta_{\scriptscriptstyle CC}\left[\ln\left(
\gamma^2\overline{\vartheta^2}\right)-1\right]=\Delta_{\scriptscriptstyle
CC}\left[\ln\left(\overline{\vartheta^2}\right)\right]
\end{equation*}
\begin{equation}
=\Delta_{\scriptscriptstyle CC}\big[\ln\left(B\right)\big]\ .
\end{equation}\\
For the latter quantity one can obtain
\begin{equation}
\Delta_{\scriptscriptstyle
CC}[\ln\left(B\right)]=\Delta_{\scriptscriptstyle CC}[B]+f(Z\alpha)=
\delta_{\scriptscriptstyle CC}[B]\ .
\end{equation}\\
The Coulomb correction becomes
\begin{equation}
\Delta_{\scriptscriptstyle CC}\left[\ln\left(
\gamma^2\overline{\vartheta^2}\right)-1\right]=\frac{\delta_{\scriptscriptstyle
CC}[B]}{\left[\ln
(\gamma^2\overline{\vartheta^2})^{\scriptscriptstyle B}-1\right]}\ .
\end{equation}\\
Taking into account (\ref{migdalparam}), we arrive at a result:
\begin{equation}\label{limcorrect2}
\delta_{\scriptscriptstyle CC}\left[\left\langle
\frac{dI}{d\omega}\right\rangle\right]= \frac{f(\xi)}{\left[\ln
(\gamma^2\overline{\vartheta^2})^{\scriptscriptstyle
B}-1\right]\bigg(1-B^{\scriptscriptstyle B}\bigg)}\ .
\end{equation}\\
The numerical values of these corrections are presented below.\\

\begin{center} {\bf Table 5.}
The relative Coulomb correction $\delta_{\scriptscriptstyle
CC}\big[\left\langle dI/d\omega\right\rangle \big]$ to the
asymptotes of the Born spectral radiation rate over the range
$\omega<\omega_{c}$ for $\beta=1$, $B^{\scriptscriptstyle B}\approx
8.46$, and
$\left(\gamma^2\overline{\vartheta^2}\right)^{\scriptscriptstyle
B}\approx 7.61$ \cite{Fomin}.
\end{center}

\begin{center}
\begin{tabular}{lcccc}
\hline \\[-3mm]
Target&Z&$\gamma^2\overline{\vartheta^2}$&$-\delta_{\scriptscriptstyle
CC}\big[\left\langle dI/d\omega\right\rangle\big]$&
$R_{\scriptscriptstyle CC}
$\\[.2cm]
\hline\\[-3mm]
Au&79&$\gamma^2\overline{\vartheta^2}\ll 1$&$0.042$&0.958\\
Au&79&$\gamma^2\overline{\vartheta^2}\gg 1$&$0.040$&0.960\\[.2cm]
\hline
\end{tabular}
\end{center}

\vspace{.25cm}

The second asymptote  is not reached \cite{Fomin} in the conditions
of experiment \cite{C-1.1.1, C-1.1.2}. Therefore we will also
consider an another limiting case corresponding to these experimental
conditions and taking into account the second term of the Moli\`{e}re
distribution function expansion (\ref{power}).

Inserting the second-order expression (\ref{2order}) for the
distribution function  into (\ref{averaged}) and integrating its
second term (\ref{W_1}), we can arrive at the following expression
for the electron radiation  spectrum at
$\mu^2=\gamma^2\overline{\vartheta^2}\gg 1$ \cite{Fomin}:

\begin{equation}\label{comleteasymp}
\left\langle \!\frac{dI}{d\omega}\!\right\rangle
\!=\!\frac{2e^2}{\pi} \!\!\left[\ln \left(\mu^2
\right)\!-\!C_{\scriptscriptstyle
E}\!\left(\!1\!+\!\frac{2}{\mu^2}\!\right)\!\!+\!\frac{2}{\mu^2}\!+\!\frac{C_{\scriptscriptstyle
E}}{B}\!-\!1\right].
\end{equation}\\

In order to obtain the Coulomb correction to the Born spectral
radiation rate from (\ref{comleteasymp}), we first calculate its
numerical value at $(\mu^2)^{\scriptscriptstyle B}\approx 7.61$ and
$B^{\scriptscriptstyle B}\approx 8.46$. Then we become $\left\langle
dI/d\omega\right\rangle^{\scriptscriptstyle B}=0.00542$. The
Bethe--Heitler formula in the Born approximation gets $\left\langle
dI/d\omega\right\rangle_{\scriptscriptstyle BH}^{\scriptscriptstyle
B}=0.00954$.

Now we calculate the numerical values of $B$ and $\mu^2$ parameters
including the Coulomb corrections. From
\begin{equation}\label{BN}
\Delta_{\scriptscriptstyle
CC}[B]=\frac{f(\xi)}{1/B^{\scriptscriptstyle B}-1} =-0.355\ ,
\end{equation}\\
we become $B=8.105$ for $Z=79$ and $B^{\scriptscriptstyle B}\approx
8.46$. The equality\\
\begin{equation*}\label{mu}
\Delta_{\scriptscriptstyle CC}\left[\ln \mu^2\right]
=\Delta_{\scriptscriptstyle CC}\left[\ln
B\right]=\Delta_{\scriptscriptstyle
CC}[B]+f(\xi)
\end{equation*}
\begin{equation}\label{mu}
=\delta_{\scriptscriptstyle CC}[B] =-0.042
\end{equation}\\
gets $\ln \mu^2=1.987$ and $\mu^2=7.295$. Inserting these values
into (\ref{comleteasymp}) we have $\left\langle
dI/d\omega\right\rangle =0.00531$. The relative Coulomb corrections
to these parameters are presented in Table 6. These corrections are
not large. Their sizes are between two to four percent, i.e. of
order of
the systematic error in the experiment \cite{C-1.1.1}.\\

\begin{center} {\bf Table 6.}
The relative Coulomb corrections in the analogue of the LPM effect
theory for $0.07\,L_{\scriptscriptstyle R}$ gold target,
$\omega<\omega_{c}$, and $\beta=1$.
\end{center}
\begin{center}
\begin{tabular}{cccc}
\hline \\[-3mm]
 $\delta_{\scriptscriptstyle CC}\left[\ln\mu^2 \right] $
&$\delta_{\scriptscriptstyle
CC}\left[\left(\frac{dI}{d\omega}\right)_{\scriptscriptstyle
0}\right]$ &$\delta_{\scriptscriptstyle
CC}\left[\left\langle\frac{dI}{d\omega}\right\rangle\right]$
&$\delta_{\scriptscriptstyle CC}\left[\Phi(s)\right]$\\[.2cm]
\hline\\[-3mm]
$-0.021$&$-0.042$&$-0.020$&$-0.021$\\[.2cm]
\hline
\end{tabular}
\end{center}

\vspace{.5cm}

Accounting the relative Coulomb correction to the Bethe--Heitler
spectrum of bremsstrahlung we find
$\left(dI/d\omega\right)_{\scriptscriptstyle BH} =0.00916$. So we
get\footnote{The obtained magnitude $\Phi(s)=0.580$ corresponds to
the value $s\sim 0.15$ (see Fig. 1 in \cite{C-4.5}).}
\begin{equation}\label{migdnum}
\left\langle\frac{dI}{d\omega}\right\rangle
=0.580\left(\frac{dI}{d\omega}\right)_{\scriptscriptstyle BH}\ .
\end{equation}\\
This leads to the value of the spectral radiation rate in terms of
$dN/[d(\log \omega)]$ $\times 1/L_{\scriptscriptstyle R}$, where $N$
is the number of events per photon energy bin per incident electron,
 $dN/[d(\log \omega)/L_{\scriptscriptstyle
R}]=0.118\times 0.580=0.068$, which agrees very well with the
experimental result over the frequency range $\omega<30$ MeV for 25
GeV and $0.7\%L_{\scriptscriptstyle \mathrm{R}}$ gold target.

This
result additionally improves the agreement between the theory
\cite{Shul'ga,Fomin} and experiment \cite{C-1.1.1,C-1.1.2}
and coincides with the result of \cite{Cont} obtained in the
eikonal approximation  (see Fig. 20a in \cite{Klein3}).\\

\section{\Large{Summary and conclusions}}

\begin{enumerate}

\item We have calculated the Coulomb corrections
($\Delta_{\scriptscriptstyle CC}\left[b \right]$,
$\Delta_{\scriptscriptstyle CC}\left[B \right]$,
$\Delta_{\scriptscriptstyle CC}\left[\ln B \right]$,
$\Delta_{\scriptscriptstyle CC}\left[\overline{\vartheta^2}\right]$,
$\Delta_{\scriptscriptstyle
CC}\left[\ln\left(\overline{\vartheta^2}\right)\right]$) and
relative Coulomb corrections ($\delta_{\scriptscriptstyle
CC}\left[\overline{\vartheta^2}\right]$, $\delta_{\scriptscriptstyle
CC}\left[B\right]$) to some important parameters of the Moli\`{e}re
multiple scattering theory for high Z  targets of experiment
\cite{C-1.1.1,C-1.1.2}, and we have showed that the corrections
$-\Delta_{\scriptscriptstyle CC}\left[b \right]$,
$-\Delta_{\scriptscriptstyle CC}\left[B \right]$ have large values
that increase up to $0.40-0.45$ for $Z=92$.

\item Using these corrections we have obtained the analytical
results for the Coulomb corrections
($\Delta_{\scriptscriptstyle
CC}\left[(dI/d\omega)_0\right]$, $\Delta_{\scriptscriptstyle
CC}\left[q\right]$, $\Delta_{\scriptscriptstyle
CC}\left[s^2\right]$, $\Delta_{\scriptscriptstyle
CC}\left[s\right]$,
 $\Delta_{\scriptscriptstyle CC}\left[s^4\right]$,
$\Delta_{\scriptscriptstyle CC}\left[\Phi(s)\right]$,
$\Delta_{\scriptscriptstyle CC}\left[\langle dI/d\omega\rangle
\right]$)  and relative Coulomb corrections
($\delta_{\scriptscriptstyle CC}\left[(dI/d\omega)_0\right]$,
$\delta_{\scriptscriptstyle CC}\left[q\right]$,
$\delta_{\scriptscriptstyle CC}\left[\Phi(s) \right]$,
$\delta_{\scriptscriptstyle CC}\left[s \right]$, $\delta_{\scriptscriptstyle CC}\left[s^2\right]$, $\delta_{\scriptscriptstyle CC}\left[s^4\right]$,
$\delta_{\scriptscriptstyle CC}\left[\langle dI/d\omega\rangle \right]$)
to the quantities of the classical Migdal LPM theory in regimes of the large
and the small LPM suppression.

\item
We have performed the calculations for the regime of small LPM
suppression over the range $1\leq s \leq \infty$,  and we have found that
the Coulomb corrections $\delta_{\scriptscriptstyle CC}\left[\langle
dI/d\omega\rangle\right]=-4.50\pm 0.05\%$ ($Z=82$) and
$\delta_{\scriptscriptstyle CC}\left[\langle
dI/d\omega\rangle\right]=-5.35\pm 0.06\%$ ($Z=92$) coincides with
the sizes of the normalization corrections $-4.5\pm 0.2\%$ for
$2\%L_{\scriptscriptstyle R}$ lead target and $-5.6\pm 0.3\%$ for
$3\%L_{\scriptscriptstyle R}$ uranium target, respectively, within
the experimental error.

\item The average $\delta_{\scriptscriptstyle CC}\left[\langle
dI/d\omega\rangle\right]$ value $\bar\delta_{\scriptscriptstyle
CC}\left[\langle dI/d\omega\rangle\right]=-4.70\pm 0.49\%$
 excellent agrees with the weighted average
 $-4.7\pm 2\%$ of the normalization correction obtained
 for 25 GeV data in the experiment \cite{C-1.1.2} in the regime of
 small LPM suppression.

\item
Thus, we managed to show that the discussed discrepancy between
theory and experiment \cite{C-1.1.1,C-1.1.2} over the range $20<\omega<500$ MeV
can be explained at least for heavy  target elements on the basis of the obtained
Coulomb corrections to the Born bremsstrahlung rate
within the Migdal LPM effect theory.

\item Finally, we found the numerical values of the relative
Coulomb corrections $\delta_{\scriptscriptstyle
CC}\left[(dI/d\omega)_0\right]$, $\delta_{\scriptscriptstyle
CC}\left[\Phi(s) \right]$, and  $\delta_{\scriptscriptstyle
CC}\left[\langle dI/d\omega\rangle \right]$ in the LPM effect theory
analogue for thin targets over the range $5<\omega<30$,  and we
demonstrated that these corrections additionally improve the
agreement between the theory \cite{Shul'ga,Fomin} and experiment
\cite{C-1.1.1,C-1.1.2}.

\item The approach based on application of the improved Moli\`{e}re
multiple scattering theory  can be useful for the analysis of
electromagnetic processes in strong crystalline fields at high
energies, the description of cosmic-ray experiments, high-energy
experiments with nuclear targets, etc.

\end{enumerate}

\section*{{Acknowledgements}}

We would like to thank Spencer Klein (Lawrence Berkeley National
Laboratory, USA) for valuable information and illuminating
discussion.

\noindent\hrulefill

\end{document}